%% file: Paper.tex
\documentclass[iop]{emulateapj}

\usepackage{graphicx}
\usepackage{apjfonts}
\usepackage{bm}

\newcommand{\domm}[1]{\ifmmode #1\else$#1$\fi}

\newcommand{\azero}{\domm{a_0}}
\newcommand{\pzero}{\domm{P_0}}
\newcommand{\Msun}{\domm{M_{\odot}}}

\newcommand{\Mch}{\domm{M_{\rm Ch}}}

\newcommand{\gcc}{\domm{{\rm g\,cm}^{-3}}}

\newcommand{\mrm}{\mathrm}

\newcommand{\arepo}{\textsc{Arepo}}
\newcommand{\gasoline}{\textsc{Gasoline}}

\newcommand{\Bmag}{\domm{|\boldsymbol{B}|}}
\newcommand{\EB}{\domm{E_B}}
\newcommand{\EBphiEB}{\domm{E_{B\phi}/E_B}}
\newcommand{\eberot}{\domm{e_B/e_\mathrm{rot}}}
\newcommand{\Zhuprep}{(C. Zhu et al. 2015, in preparation)}
\newcommand{\Lsun}{\domm{L_\odot}}

\newcommand{\eqbegin}{\begin{equation}}
\newcommand{\eqend}{\end{equation}}

\defcitealias{vkercj10}{vKCJ10}
\defcitealias{loreig09}{LIG09}
\defcitealias{loreig09}{LIG09}
\defcitealias{zhu+13}{Z13}
\defcitealias{ji+13}{J13}
\newcommand{\citeal}{\citetalias}

\begin{document}

\title{Magnetized Moving Mesh Merger of a Carbon-Oxygen White Dwarf Binary}

\author{Chenchong Zhu, R\"{u}diger Pakmor, Marten H. van Kerkwijk, Philip Chang}
\email{cczhu@astro.utoronto.ca}

\date{\today}

\input{Abstract.tex}

\maketitle

\input{Introduction.tex}
\input{Methods.tex}
\input{Results.tex}
\input{Discussion.tex}

\bibliography{AutoBibliography}

\end{document}

%% file: Abstract.tex
\begin{abstract}
White dwarf (WD) binary mergers are possible progenitors to a number of unusual stars and transient phenomena, including type Ia supernovae.  To date, simulations of mergers have not included magnetic fields, even though they are believed to play a significant role in the evolution of the merger remnant.
We simulated a 0.625 - 0.65 {\Msun} carbon-oxygen WD binary merger in the magnetohydrodynamic moving mesh code {\arepo}.  Each WD was given an initial dipole field with a surface value of $\sim10^3$ G.
As in simulations of merging double neutron star binaries, we find exponential field growth within Kelvin-Helmholtz instability-generated vortices during the coalescence of the two stars.  The final field has complex geometry, and a strength $>10^{10}$ G at the center of the merger remnant.  Its energy is $\sim2\times10^{47}$ ergs, $\sim0.2$\% of the remnant's total energy.
The strong field likely influences further evolution of the merger remnant by providing a mechanism for angular momentum transfer and additional heating, potentially helping to ignite carbon fusion.

\end{abstract}

%% file: Introduction.tex
\section{Introduction}
\label{sec:intro}

White dwarf (WD) binaries are common end products of binary stellar evolution.  Gravitational wave emission, magnetic braking or the influence of a third body will cause a fraction of these to merge, producing a diversity of unusual stars and electromagnetic transients.  In particular, for double carbon-oxygen (CO) WD mergers, the final outcome could be a massive and rapidly rotating WD (eg. \citealt{segrcm97}), an accretion-induced collapse into a neutron star (NS) \citep{saion85}, or a nuclear explosion that might resemble a type Ia supernova (SN Ia).  Understanding the conditions that lead to each outcome requires understanding the merging process, which has, in the last decade, been investigated with increasingly sophisticated 3D hydrodynamic simulations \citep{loreig09, pakm+10, dan+12, dan+14, rask+12, zhu+13, moll+14}.  However, one fundamental piece missing in WD merger studies so far is magnetic fields.

Mergers (that do not immediately explode) are expected to produce differentially rotating merged objects - ``merger remnants'' - that are susceptible to magnetic dynamo processes such as the magnetorotational instability (MRI; \citealt{balbh91}), Tayler-Spruit dynamo (e.g. \citealt{spru02}), and the $\alpha\omega$ dynamo (if convection occurs in the inner disk; \citealt{garc+12}).  It has therefore long been suspected that they can generate strong fields, and recent 2D simulations of MRI in the remnant \citep{ji+13} have indeed shown amplification of a weak seed field to $>10^{10}$ G.  Magnetic shear from these fields transports angular momentum over a timescale of $\sim10^4 - 10^8$ s \citep{vkercj10, shen+12} -- far shorter than the thermal timescale of the remnant -- and also (non-locally) heats the remnant.  The latter, combined with loss of rotational support from angular momentum transport, could push remnant temperatures past the point of carbon ignition ($\sim6\times10^8$ K for densities between $10^5 - 10^7$ \gcc), leading to either stable nuclear burning or a runaway.  This mechanism could potentially drive nuclear runaways even in remnants with masses below the Chandrasekhar Mass \Mch\ that have traditionally been considered stable \citep{vkercj10}.

While field growth after the merger has been explored, field growth \textit{during} the merger is also expected, and can have a profound impact on the post-merger magnetic evolution.  Magnetohydrodynamic (MHD) double NS binary merger simulations (eg. \citealt{pricr06, giac+14, kiuc+14}) have found that Kelvin-Helmholtz vortices produced along the shear interface between the coalescing stars can amplify field strengths by orders of magnitude \citep{oberam10, zrakm13}.  The same should hold true for WD mergers.  Motivated by this, we present the first MHD simulation of a CO WD binary merger.



%% file: Methods.tex
\section{Methods} 
\label{sec:codes}

We employ the moving-mesh code \arepo\ \citep{spri10}, which solves the equations of ideal MHD on a Voronoi mesh coupled with self-gravity.  We operate the code in its pseudo-Lagrangian mode, so that the mesh-generating points that define the Voronoi grid move with the local velocity of the fluid.  To conserve angular momentum to within $\sim2$\% of its initial value, we use the latest improvements to time integration and gradient estimate \citep{pakm+15}.  \arepo's MHD implementation is described in \cite{pakmbv11} and \cite{pakmv13}; we use the \cite{powe+99} eight-wave scheme for divergence control.  Our simulation ignores outer hydrogen and helium layers, composition gradients, and nuclear reactions (negligible for sub-\Mch\ CO WD mergers; \citealt{loreig09,rask+12}).

We model the merger of two CO WDs with masses of 0.625 and 0.65 \Msun, respectively, in a circular, unsynchronized binary with initial separation $\azero = 2.20\times10^9$ cm (corresponding period $\pzero = 49.5$ s), chosen (using the estimate of \citealt{eggl83}) such that the lower-mass WD just fills its Roche lobe.  We chose masses typical of the narrowly peaked empirical mass distribution of field CO WDs \citep{klei+13}.  Our initial conditions are very similar to those of the 0.625 - 0.65 \Msun\ binary simulated with smoothed particle hydrodynamics (SPH) in \citeauthor{zhu+13} (\citeyear{zhu+13}; henceforth \citeal{zhu+13}).  As in \citeal{zhu+13}, both WDs are generated with a uniform initial temperature of $5\times10^6$ K (the corresponding thermal pressure is dynamically irrelevant) and a uniform composition of equal parts carbon and oxygen by mass.  They are separately relaxed to hydrostatic equilibrium using the SPH code \gasoline\ \citep{wadssq04} and added to \arepo\ by converting the SPH particles to mesh-generating points while retaining their conservative quantities (mass, momentum and energy).  A uniform $10^{-5}$ \gcc\ background grid fills up a $10^{12}$ cm box centered on the binary. Each WD is given a (dynamically irrelevant) dipole seed magnetic field with an equatorial surface value of $10^3$ G (and corresponding central field of $\sim2\times10^7$ G). The fields are overlapped when the two stars are placed into a binary.

The mass resolution of our simulation is $m_\mathrm{cell} \approx 1\times10^{-6}$ \Msun.  We utilize explicit refinement and derefinement \citep{voge+12} to keep cell masses within a factor of two of $10^{-6}$ \Msun\ and to ensure adjacent cells differ by less than a factor of 10 in volume.


%% file: Results.tex
\section{Results}
\label{sec:results}

\begin{figure}
\centering
\includegraphics[angle=0,width=1.0\columnwidth]{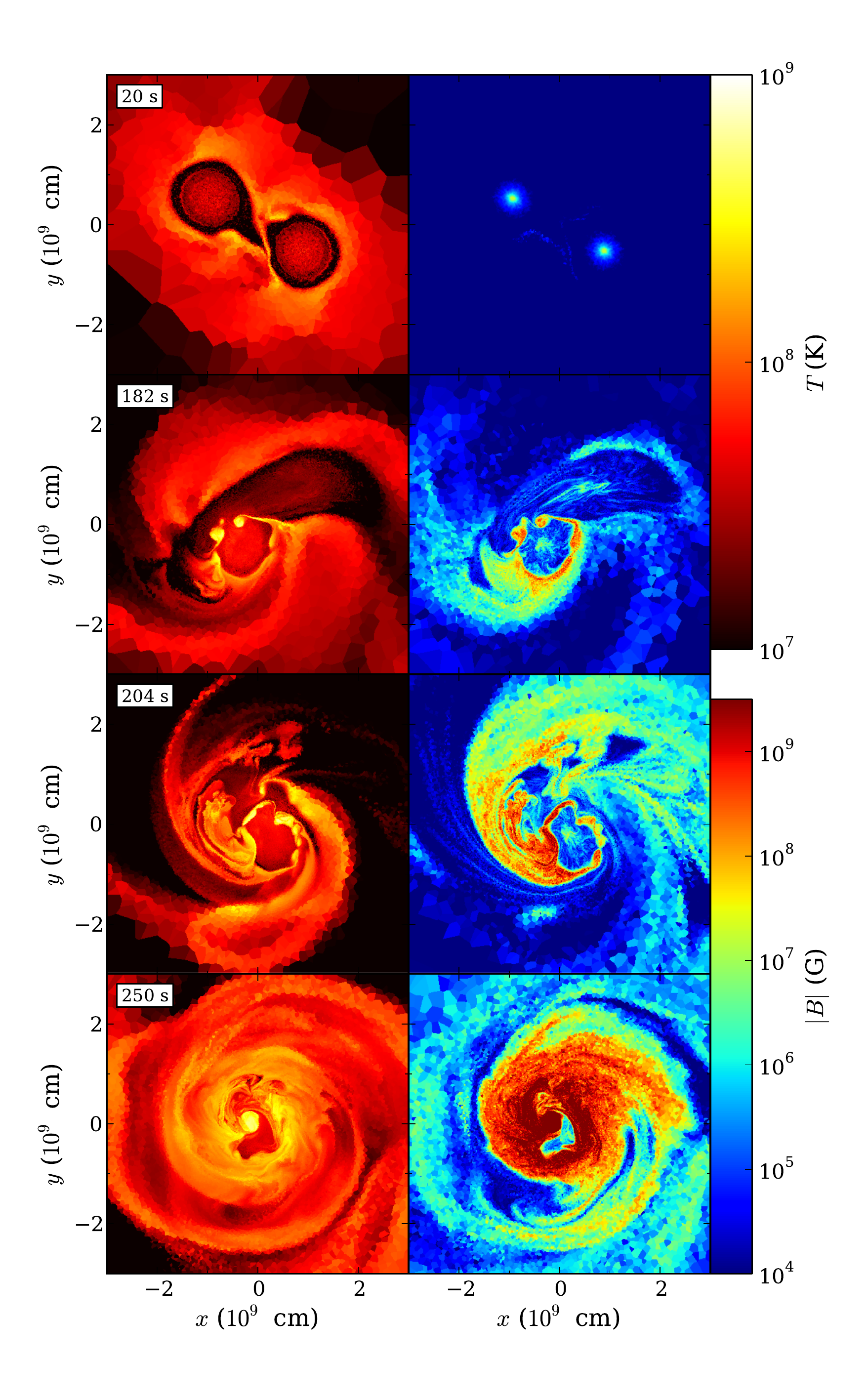}
\caption{Series of temperature $T$ (left column) and magnetic field strength \Bmag\ (right) logarithmic intensity profiles in the equatorial plane of the merger for four snapshots in time (rows; time indicated at the top left of each row).}
\label{fig:snapshots}
\end{figure}

\begin{figure}
\centering
\includegraphics[angle=0,width=1.0\columnwidth]{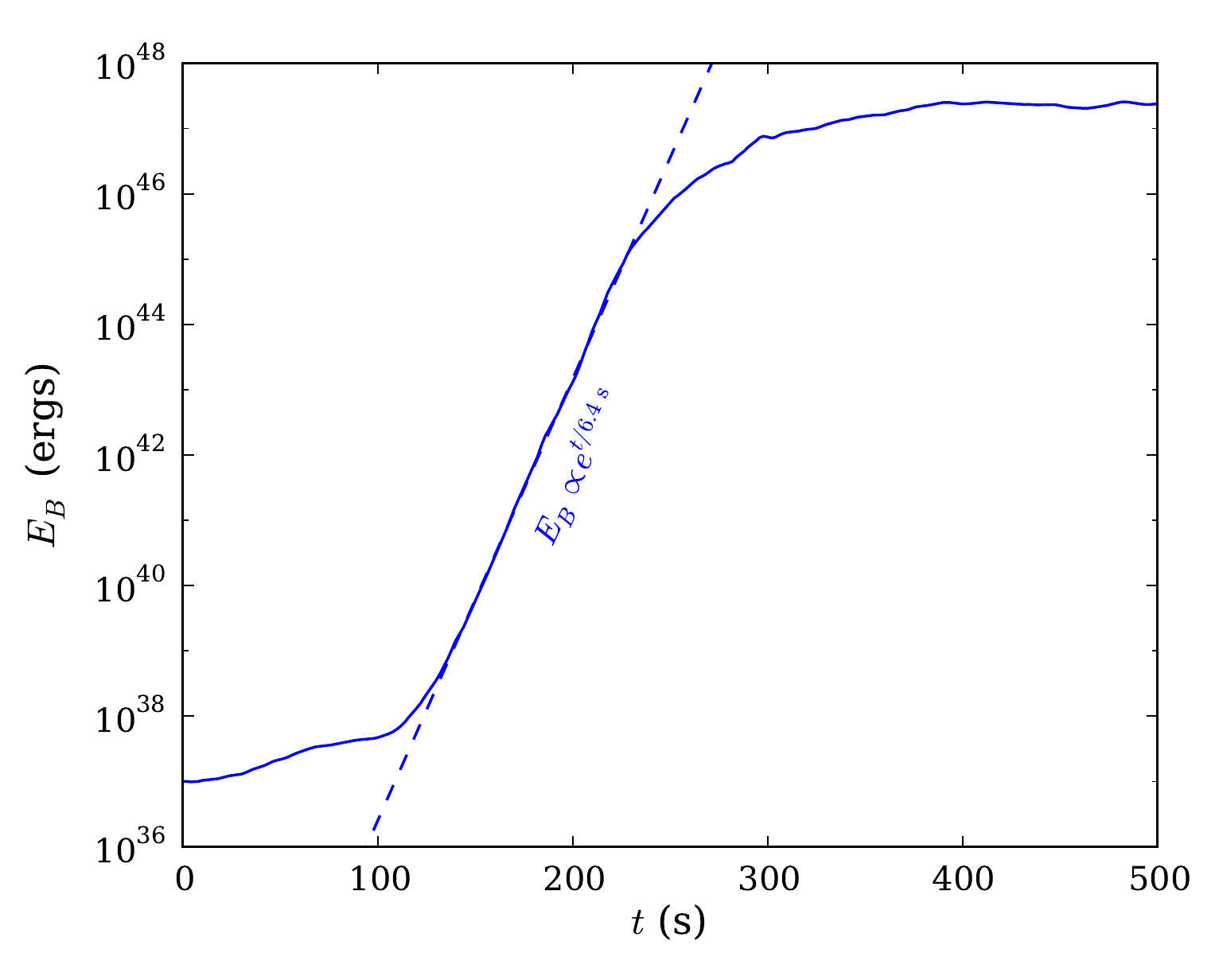}
\caption{Total magnetic energy \EB\ over time, with a best fit to the rapid exponential growth (dashed $\EB \propto e^{t/6.4\,\mrm{s}}$ line).}
\label{fig:bgrowth}
\end{figure}

We depict the evolution of the binary in Fig. \ref{fig:snapshots}, highlighting temperature $T$ and magnetic field strength \Bmag.  Fig. \ref{fig:bgrowth} shows the growth of total magnetic energy \EB\ over time.

In the first stage of the merger, up to $\sim180$ s, the 0.625 \Msun\ donor WD transfers mass to the 0.65 \Msun\ accretor for about 3.5 orbits before fully disrupting.  Because our initial conditions are approximate -- the WDs are not initially tidally deformed -- mass transfer begins in spurts as the WDs stretch in response to the binary potential, and occurs at a rate that is artificially high \citep{dan+11}.  The early mass transfer shears the atmospheres of both WDs.  As a result, \EB\ grows roughly linearly in the first $\sim100$ s, reaching about quadruple its initial value.  Since the initial mass transfer rate is artificially high, this growth is likely an overestimate, but remains negligible compared to what follows.


By $\sim120$ s, mass transfer becomes steady, and a stream of material from the donor wraps around the accretor, forming a shear layer.  Along it, the Kelvin-Helmholtz instability generates a string of hot vortices that exponentially amplify their entrained magnetic fields.  This is illustrated in Fig. \ref{fig:snapshots} (row 2), where the hot vortices along the donor-accretor interface correspond to regions of high field strength.  At $\sim180$ s, tidal forces between the two WDs become strong enough to fully disrupt the donor, which then coalesces with the accretor over $\sim50$ s.  During this time, infalling donor material spirals into the accretor, severely deforming the accretor while carrying the string of magnetized vortices toward the system's center of mass (CM).  Fig. \ref{fig:bgrowth} shows \EB\ growing exponentially by a factor of $\sim10^7$ over $\sim100$ s, with an $e$-folding time $\tau = 6.4$ s, comparable to the typical turnover timescale of the largest eddies $2\pi R_\mrm{eddy}/\Delta v_\mrm{shear} \sim 3\,\mrm{s}$, where $\Delta v_\mrm{shear}$ is the velocity difference across the shear layer.


By $\sim250$ s many of the vortices have merged together into a hot, rapidly rotating underdense void at the CM (Fig. \ref{fig:snapshots}, row 4).  Magnetic growth within the void begins to saturate as its magnetic and kinetic energy approach equipartition.  The rate of field growth slows as well, with \EB\ growing another two orders of magnitude over $\sim150$ s before plateauing at $\sim2\times10^{47}$ ergs at $\sim400$ s.


\begin{figure*}
\centering
\includegraphics[angle=0,width=2.0\columnwidth]{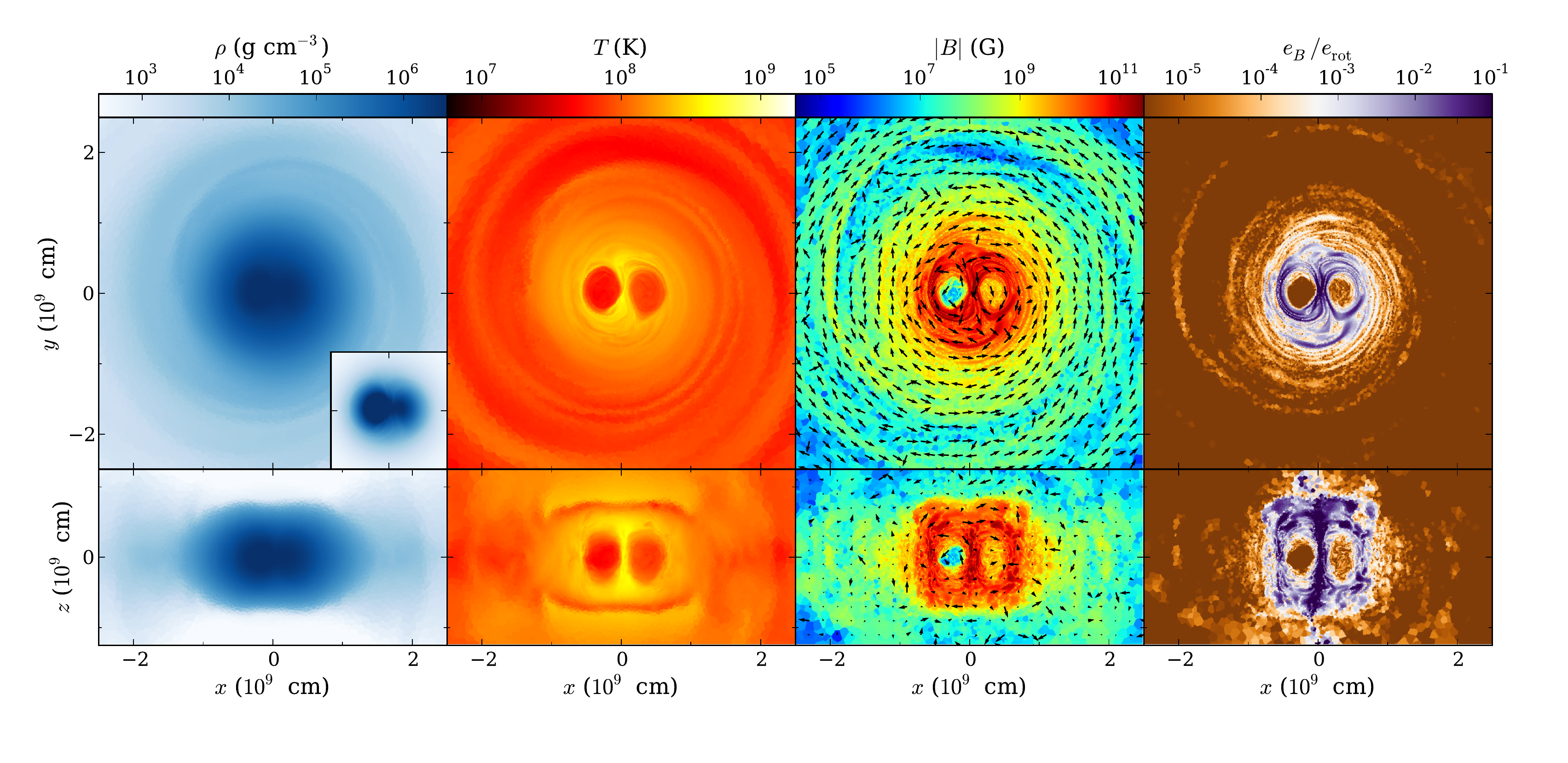}
\caption{From leftmost to rightmost column, equatorial plane (top row) and polar (bottom) logarithmic intensity profiles of density $\rho$, temperature $T$, magnetic field strength \Bmag\ and ratio of magnetic to rotational energy density \eberot\ for the simulation at 400 s ($\sim170$ s after coalescence).  The equatorial plane density plot includes a linear profile of the remnant core (with the same $x$ and $y$ scale as the logarithmic profile) to show its shape.  Arrows in the magnetic field strength plots indicate field directions, with their lengths equal to the fraction of the field that lies along the $xy$ plane (top frame) and $xz$ plane (bottom).}
\label{fig:remnant}
\end{figure*}


In Fig. \ref{fig:remnant} we show the density $\rho$, $T$, \Bmag\ and ratio of magnetic to rotational energy density \eberot\ of the merger remnant at 400 s.  The remnant consists of a dense, degeneracy-supported core containing $\sim60$\% of the remnant's mass, a partly thermally supported hot envelope that surrounds the core, and a rotationally supported disk, a configuration similar to the SPH 0.625-0.65 \Msun\ remnant from \citeal{zhu+13}.  The \arepo\ remnant core, however, has distinctly non-axisymmetric density and temperature structures, unlike the SPH simulation which achieves axisymmetry $\sim170$ s after coalescence (\citeal{zhu+13}, online figure set Fig. 1.16).  The magnetic fields are too weak during the merger to have an effect on merger dynamics, so these contrasts are due to differences in the \textit{hydrodynamic} schemes between \arepo\ and SPH \Zhuprep.



The remnant magnetic field configuration is complex: while field lines are coherent along ``strands'' of high field strength, neighboring strands often point in opposite directions (see Fig. \ref{fig:remnant}).  In the core, the volume-averaged field strength is $4\times10^{10}$ G, but strands of $>10^{11}$ G field perforate the core.  The field remains $>10^9$ G near the core-disk interface at $\sim 10^9$ cm, before dropping below $10^7$ G at $\gtrsim 3\times10^9$ cm.  The total magnetic field energy is $\sim0.2$\% the total, $\sim0.6$\% the total rotational, and $\sim6$\% the total differential rotation energy of the remnant.\footnote{Differential rotation energy of a cell is estimated with $E_\mrm{drot} = m_\mrm{cell}|\boldsymbol{v}||\nabla\times\boldsymbol{v}|V_\mrm{cell}^{1/3}$.}  This energy is roughly equally partitioned into toroidal and poloidal field components, with the ratio of poloidal to total magnetic energy $\EBphiEB = 0.62$.  Studies of local field amplification within Kelvin-Helmholtz vortices predict magnetic growth saturates when the magnetic and kinetic energy densities are close to equipartition \citep{oberam10, zrakm13}.  In our simulation, this is only the case for the strands of $>10^{11}$ G field, where magnetic energy density is $\sim7$\% ($\sim47$\%) the rotational (differential rotation) energy density (see Fig. \ref{fig:remnant}, column 4).  It is possible that because the strands are distributed throughout the core, they drive the core's overall evolution and inhibit further magnetic amplification in their surroundings.


Some of the magnetized accretion stream is ejected during coalescence and integrates into the inner disk ($\sim1 - 3\times10^9$ cm), producing a $10^7 - 10^8$ G field by 400 s.  This field has a negligible hydrodynamic effect on the disk (magnetic energy density to pressure ratio $e_\mrm{B}/P \approx 3\times10^{-5}$ at $2\times10^9$ cm), and, unlike the field in the core, has \textit{not} saturated: \Bmag\ continues to grow exponentially with $\tau \sim 200$ s.\footnote{The remnant disk is generally poorly resolved, even at the highest resolution used in Sec. \ref{ssec:restest}; this may artificially slow the disk field growth rate.}

%% file: Discussion.tex
\section{Robustness Tests}
\label{sec:robust}

\begin{figure}
\centering
\includegraphics[angle=0,width=1.0\columnwidth]{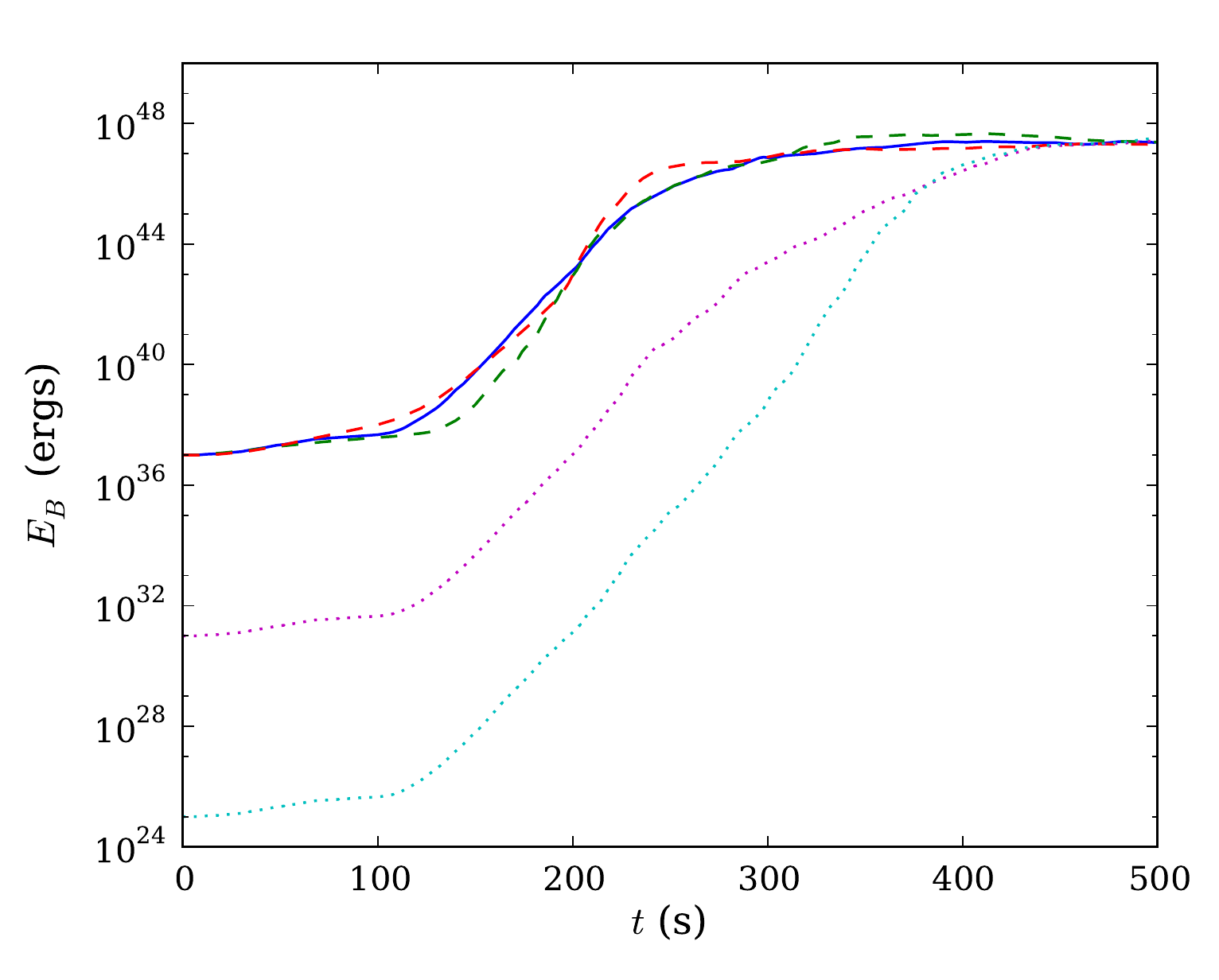}
\caption{Total magnetic energy \EB\ over time for the fiducial (solid blue; $m_\mathrm{cell} \approx 1\times10^{-6}$ \Msun, equatorial surface field strength $10^3$ G) simulation and the robustness tests.  Dashed lines represent low (green; $m_\mathrm{cell} \approx 5\times10^{-6}$ \Msun) and high resolution (red; $m_\mathrm{cell} \approx 2\times10^{-7}$ \Msun) simulations.  Dotted lines represent $1$ G (magenta) and $10^{-3}$ G (cyan) low initial field simulations.}
\label{fig:bcomp}
\end{figure}

\subsection{Resolution Test}
\label{ssec:restest}

To check that our results are not resolution-limited, we performed simulations, otherwise identical to the fiducial one in Sec. \ref{sec:results}, with lower and higher mass resolutions of $m_\mathrm{cell} \approx 5\times10^{-6}$ \Msun\ and $m_\mathrm{cell} \approx 2\times10^{-7}$ \Msun, respectively.  Fig. \ref{fig:bcomp} compares the \EB\ evolution between these simulations (dashed lines) and the fiducial one (solid).

The three runs are qualitatively identical.  Donor disruption and the start of exponential field growth occurs $\sim0.75$ orbital periods earlier at low resolution, and $\sim0.5$ periods later at high, because the outer layers of the WDs are better captured and the differences in hydrostatic equilibrium between \arepo\ and \gasoline\ are less pronounced at higher resolution.  Exponential growth rates are similar between the runs - the \EB\ $e$-folding time is $\tau = 4.9$ s for the low resolution run, faster than $\tau = 6.4$ s for the fiducial.  At high resolution, the growth curve appears to be separated into two phases, with $\tau = 7.8$ s before coalescence, and $\tau = 3.9$ s during it.  The total magnetic energy at the end of exponential growth is also similar - at 400 s, \EB\ is $\sim4\times10^{47}$ ergs in the low resolution run and $\sim1.5\times10^{47}$ ergs in the high, compared to $\sim2\times10^{47}$ ergs at the fiducial resolution.  The fiducial and high resolution runs also qualitatively have very similar magnetic field structures by 400 s.  Our fiducial resolution of $1\times10^{-6}$ \Msun\ therefore appears sufficient for qualitatively capturing the growth and final field configuration of the merger.

In their MHD disk galaxy simulations, \cite{pakmv13} find faster field growth and higher field strength at saturation in their lowest resolution run, which they attribute to larger divergence errors at lower resolution.  We perform a similar test, and also see a trend of decreasing divergence error (and more accurate magnetic evolution) at higher resolution, though the errors of all our simulations are at least a factor of two smaller than any reported in \cite{pakmv13}.  The errors are highly localized in space and trace steep magnetic gradients, suggesting they contribute only to small-scale variations in the magnetic field.


\subsection{Changing the Seed Field Strength}
\label{ssec:seed}

To understand the dependence of our results on the initial seed field, we ran two additional simulations in which we decreased the strength of the seed field by $3$ and $6$ orders of magnitude leading to an initial equatorial surface field of $1$ G (central field $\sim2\times10^{4}$ G), and $10^{-3}$ G ($\sim20$ G), respectively.  Fig. \ref{fig:bcomp} shows their \EB\ evolution (dotted lines).  We find the growth curves to be homologous between both low initial field runs and the fiducial one -- differing only by the ratios of seed \EB\ -- up to $\sim200$ s, with the $e$-folding time for exponential amplification approximately $6.5$ s for all three runs.  By $\sim250$ s, the field in the fiducial simulation begins to plateau, while amplification (of initially weaker fields) continues for several hundred more seconds in the low initial field runs.  For both runs, \EB\ plateaus at $\sim3\times10^{47}$ ergs, comparable to the fiducial $\sim2\times10^{47}$ ergs.  Because the fields in the low initial field runs remain dynamically irrelevant for longer, however, their structures differ from that of the fiducial run and resemble more the crescent in Fig. \ref{fig:snapshots}, row 4.  The disk field does not saturate in any simulation -- its strength is proportional to the strength of the seed field, and is thus much weaker in the low initial field runs.  Our tests thus suggest that the exponential growth, growth timescale and plateau \EB\ are robust to changes in initial field strength, while the remnant field configuration is more sensitive to the choice of seed field.

\section{Discussion}
\label{sec:discussion}

We have shown that the merger of a 0.625 - 0.65 \Msun\ CO WD binary produces a strong, $>10^{10}$ G magnetic field with a complex structure that winds through the remnant core and into the inner disk.  Similar to previous simulations of binary NS mergers, the strong field is generated by dynamo action within Kelvin-Helmholtz vortices formed during the coalescence of the two WDs.  Since these vortices are ubiquitous in WD mergers, strong magnetic fields are a likely feature of all merger remnants.  The degree to which a field permeates the remnant core depends on how thoroughly the donor and accretor mix during coalescence, which itself is sensitive to initial conditions such as the degree of synchronization between the WDs, or how accurately their tidal bulges and early mass transfer are captured \citep{dan+11, dan+14}.  A parameter space study of magnetized mergers is needed to investigate the range of possible remnant field configurations.


We note that NS mergers simulated in Eulerian grid codes generally show \EB\ growing by only a factor of $\sim10^2-10^3$ during coalescence, compared to the $\sim10^9$ we see, despite these simulations having resolutions comparable or superior to our low resolution \arepo\ run.  This weaker growth is also inconsistent with the amplification to local kinetic equipartition seen in small-scale simulations \citep{oberam10, zrakm13}, and is attributed to insufficient resolution in the NS merger simulations (\citealt{giac+14,kiuc+14}, though see \citealt{dionar15}).  \cite{giac+14} incorporated a subgrid magnetic amplification model, calibrated using \cite{zrakm13}s results, into their Eulerian NS merger simulation, and found \EB\ amplification by a factor of $\sim10^{10}$ over a single dynamical time.\footnote{\cite{pricr06}s SPH simulations also show strong amplification; while runs using their Euler potential MHD method suffer exaggerated field growth from improper boundary conditions, their $\boldsymbol{B}$-based runs show similar results \citep{pric12}.}  This suggests \arepo\ may be better able to resolve small-scale velocity structures than an Eulerian code at comparable resolution, or better able to couple these structures to magnetic growth.  Further work is needed to understand the magnetic field growth in detail.

The density profile of the remnant remains non-axisymmetric for hundreds of seconds after coalescence.  As a result, the core continues to evolve dynamically, and by 400 s has begun to launch a pair of spiral waves into the surrounding medium (see the density panel of Fig. \ref{fig:remnant}), which transport angular momentum on a timescale rivalling that for magnetic shear.  While \cite{kash+15} report a similar spiral instability in their Eulerian remnant evolution simulation, SPH simulations like those of \citeal{zhu+13} rapidly achieve axisymmetry after coalescence and do not form spiral waves.  As noted earlier, this difference between \arepo\ and SPH simulations is a product of the differences between their hydrodynamic schemes.  Further study is needed to understand these differences and their consequences for remnant evolution \Zhuprep.


The post-merger evolution of the remnant has been followed to $\sim3\times10^4$ s with axisymmetric cylindrical (two-dimensional) Eulerian grid simulations \citep{schw+12,ji+13}.  As described earlier, \cite{ji+13}s MHD simulation of a 0.6 - 0.6 \Msun\ remnant shows the development of a strong magnetic field due to MRI.  The subsequent heating and angular momentum transport due to the fields pushes core temperatures to ignition, supporting the possibility of a nuclear runaway within sub-\Mch\ remnants.  Their results are, however, likely sensitive both to their initial hydrodynamic conditions -- which may have artificially high core temperatures -- and their chosen seed magnetic field, a pure poloidal one to optimize the onset of MRI.  Our much stronger poloidal-toroidal field could substantially change post-merger evolution.  Moreover, the persistence of a non-axisymmetric remnant core will lead to evolution that clearly cannot be captured in a axisymmetric cylindrical grid.  We therefore stress the need to perform high-resolution three-dimensional simulations of post-merger evolution to determine the final fate of the remnant.


There are a number of potentially observable consequences of the magnetic fields produced by the merger.  \cite{ji+13} note the creation of a magnetized corona and biconal jet in their simulations, which act in concert to cause an outflow of material near the remnant's poles.  This outflow eventually unbinds $\sim10^{-3}$ \Msun\ of material, and \cite{belo14} estimates it should lead to an optical transient with a duration of $\sim1$ day and peak $L \sim10^{8}$ \Lsun, which should be detectable by optical surveys.

If a remnant later experiences a nuclear runaway and explodes as a SN Ia, its magnetic field will increase the late-time emission by trapping positrons (produced by $^{56}$Co $\beta^+$ decay) that would otherwise escape the ejecta.  The trapping efficiency depends on the strength and configuration of the remnant magnetic field, with a locally entangled $\sim10^{11}$ G field -- similar to our findings -- well able to trap positrons past 1000 days \citep{ruizs98}.  This is in line with observed late-time SN Ia light curves (most recently \citealt{kerz+14}).

Those remnants that do not explode will retain strong fields when they reach quiescence, and populate the high-mass tail of the distribution of isolated high-field magnetic WDs \citep{garc+12,kule+13,wicktf14,brig+15}.  Since these remnants will have temperatures high enough to fuse any remaining hydrogen and helium they possess, their properties might eventually be akin to the recently discovered hot DQ WDs (e.g. \citealt{dufo+13}).  These WDs have carbon-dominated atmospheres and $T_\mrm{eff} \approx 2\times10^4$ K, are often strongly magnetized ($\sim1$ MG) and sometimes have monoperiodic photometric variability (possibly due to rapid rotation).  Their origins remain unclear \citep{alth+09, lawr+13, will+13}.  Dunlap \& Clemens (in press) recently found that, if most known hot DQs are massive ($M \gtrsim 0.95\Msun$), their population's velocity dispersion corresponds to a kinematic age much older than what would be inferred from their temperatures, suggesting that at least some hot DQs are WD-WD merger remnants.  If so, their observed properties would constrain merger and remnant evolutionary models, and double-degenerate channels for SNe Ia.

\vspace{5mm}

We thank Christopher Thompson, Christopher Matzner, Volker Springel, Bart Dunlap, Yuri Levin, Henk Spruit, Ue-Li Pen and Stephen Ro for their insights into magnetohydrodynamics and simulations.  This work utilized the SciNet HPC Consortium's GPC supercomputer \citep{loke+10}.  C.Z. acknowledges support from the Natural Sciences and Engineering Research Council (NSERC) Vanier Canada Graduate Scholarship.  R.P. acknowledges support by the European Research Council under ERC-StG grant EXAGAL-308037 and by the Klaus Tschira Foundation.  P.C. is supported by the NASA ATP program through NASA grant NNX13AH43G, and NSF grant AST-1255469.